\documentclass[11pt]{article}

\usepackage[]{graphicx}

\newcommand{\gfrac}[2]{\displaystyle\frac{#1}{#2}}

\newcommand{\dd}{\mbox{d}}

\title{Linear Polarimetry with $\gamma \rightarrow e^+e^-$ conversions}

\author{Denis Bernard
 \\
 LLR, Ecole Polytechnique, CNRS/IN2P3, 91128 Palaiseau, France
}

\begin{document} 

\maketitle 

\begin{center}
\large \textbf{Presented at the Polarised Emission from Astrophysical Jets Conference, June 12-16, 2017, Ierapetra, Greece }
\end{center}

\begin{abstract}$\gamma$ rays are emitted by cosmic sources by non-thermal
 processes that yield either non-polarized photons, such as those
 from $\pi^0$ decay in hadronic interactions, or linearly polarized
 photons from synchrotron radiation and the inverse-Compton
 up-shifting of these on high-energy charged particles.
Polarimetry in the MeV energy range would provide a powerful tool to
discriminate among ``leptonic'' and ``hadronic'' emission models of
blazars, for example, but no polarimeter sensitive above 1\,MeV has
ever been flown into space.
Low-$Z$ converter telescopes such as silicon detectors are developed
to improve the angular resolution and the point-like sensitivity below
100 MeV.
We have shown that in the case of a homogeneous, low-density active
target such as a gas time-projection chamber (TPC), the single-track
angular resolution is even better and is so good that in addition the
linear polarimetry of the incoming radiation can be performed.
We actually characterized the performance of a prototype of such a
telescope on beam.
Track momentum measurement in the tracker would enable
calorimeter-free, large effective area telescopes on low-mass
space missions.
An optimal unbiased momentum estimate can be obtained, in the tracker
alone, based on the momentum dependence of multiple scattering, from a
Bayesian analysis of the innovations of Kalman filters applied to the
tracks.
\end{abstract}


\section{MeV $\gamma$-ray astronomy}

$\gamma$-ray astronomy is suffering from a huge sensitivity gap
between the sub-MeV energy range for which Compton telescopes are very
efficient and the energy range above 100\,MeV for which pair
telescopes are very efficient \cite{Schonfelder}. 
From pair creation threshold (1\,MeV) to 100\,MeV, the main issue is
the difficulty to reject true-photon backgrounds due to the bad
single-photon angular resolution: The Fermi-LAT, for example, has
published results mainly above 100\,MeV \cite{McEnery:eASTROGAM}.
 
\begin{figure}[ht]
\includegraphics[width=0.475\linewidth]{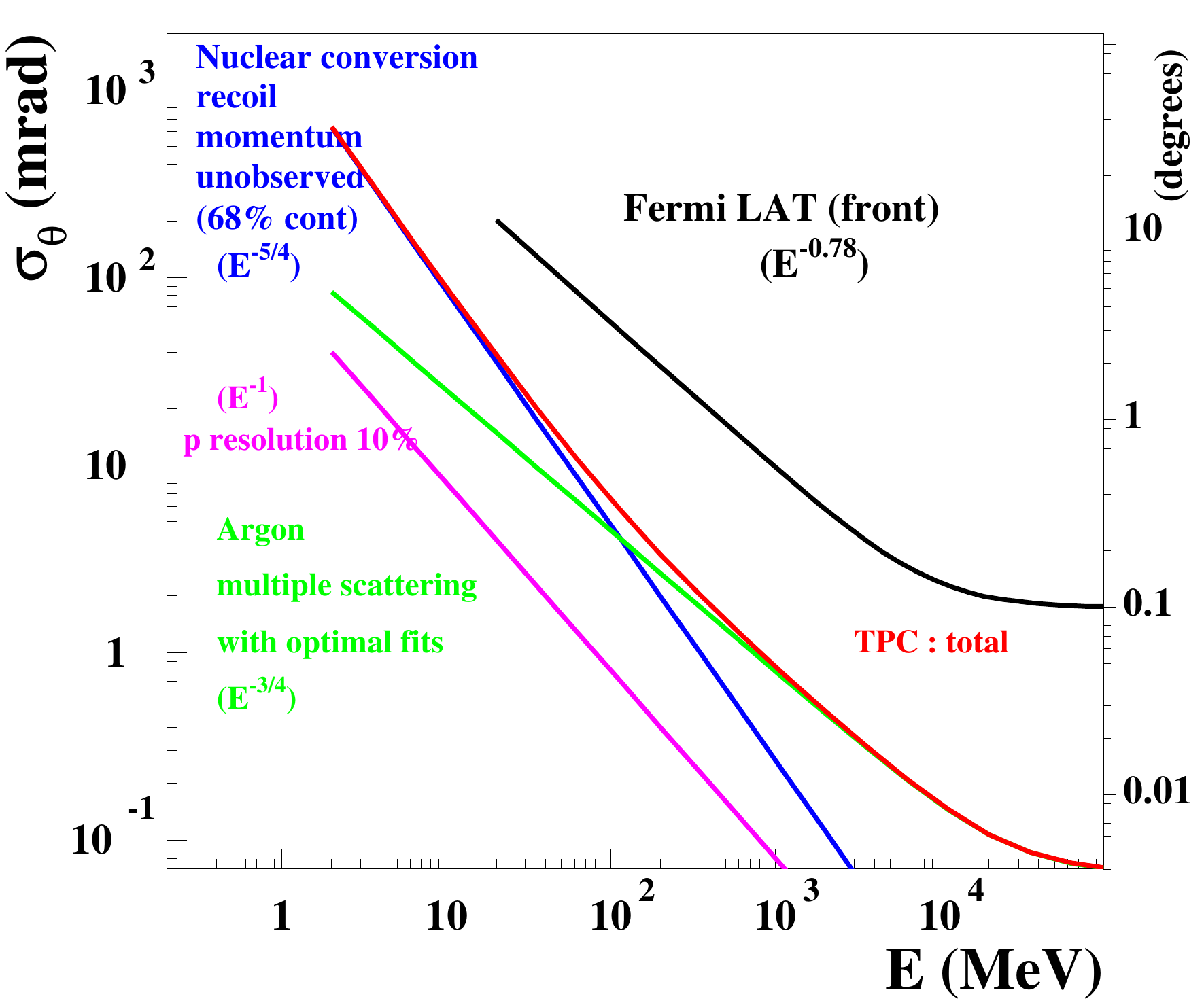}
 \hfill 
 \includegraphics[width=0.482\linewidth]{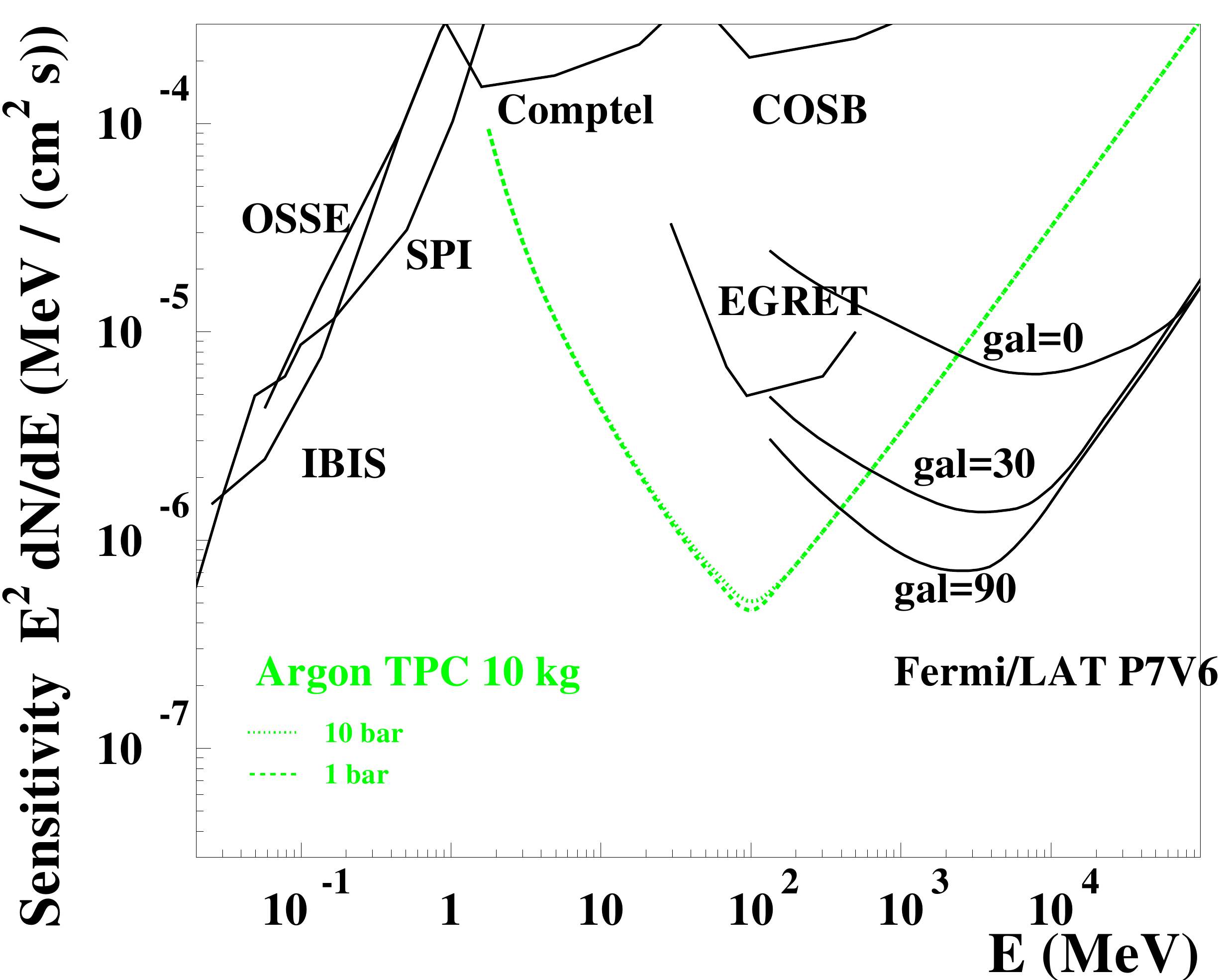}
 \caption{
Left : the three contributions to the single-photon angular
resolution, namely the lack of recoil ion momentum measurement (blue),
the single-track angular resolution (green), the single-track momentum
resolution (cyan, assumed here to be of 10\,\%) and their total (red),
as a function of the incoming photon energy, compared to that of the
Fermi-LAT.
Right : the point-like source sensitivity of a 10\,kg gas TPC
telescope, computed à la Fermi-LAT, as a function of the incoming
photon energy, compared to that of past and present telescopes.
Both adapted from \cite{Bernard:2012uf}.
\label{fig:TPC}
 }
\end{figure}

Efforts are in progress to improve on the contribution to the angular
resolution due to the multiple scattering of the electron and of the
positron in the detector.
With Tungsten-converter-free silicon wafer stacks \cite{AMEGO,E-Astrogam:2016}
or emulsion detectors \cite{Takahashi:2015jza}, an
improvement by a factor of 3 can be expected, at 100\,MeV, with
respect of the angular resolution of the Fermi-LAT \cite{Ackermann:2012kna}.
With gas detectors, up to a factor of 10 can be obtained 
(Fig. \ref{fig:TPC} left and \cite{Bernard:2012uf}).
For energies lower than 100\,MeV, the point-like source sensitivity of
gas detectors is dominated by true-photon background rejection and is
excellent, thanks to the improved angular resolution
(Fig. \ref{fig:TPC} right).
At higher energies, it is dominated by photon statistics and therefore
a gas detector (10\,kg for Fig. \ref{fig:TPC}) cannot compete with the
multi-ton Fermi-LAT.

\section{Polarimetry with pairs}

In contrast to low energy (radiowaves, optics) polarimetry that
formed the core of this conference, and for which polarimetry is
performed with detectors that measure electric fields or light
intensities, at high energies photons are observed individually: a
photon conversion in the detector is named an ``event''.
Whatever the process at work (photo-electric effect, Compton
scattering, pair conversion), due to the
$J^{PC} = 1^{--}$ quantum numbers of the photon, the one-dimensional
(1D) differential cross section takes the form:

\begin{equation}
\gfrac{\dd \sigma}{\dd \varphi} \propto 
\left(
1 + A \times P \cos(2(\varphi - \varphi_0))
\right).
\label{eq:modulation}
\end{equation}

The modulation factor of the cosine, $A \times P$, is the product of
the polarization asymmetry of the conversion process, $A$, and of the
linear polarization fraction of the incoming radiation, $P$.
$\varphi$ is the azimuthal angle of the event, that is, an angle that
measures the orientation of the event in a plane orthogonal to the
direction of propagation of the incoming photon
and
$\varphi_0$ is the polarization angle of the incoming radiation.

In the case of pair conversion, $\gamma \rightarrow e^+e^-$, the final
state is described by five variables that can be chosen to be the
azimuthal angles and the polar angles of the electron and of the
positron, respectively, and the fraction of the energy of the incident
photon that is carried away by the position, $\phi_-$, $\theta_-$,
$\phi_+$, $\theta_+$ and $x_+$.
The full (5D) unpolarized differential cross section has been obtained
by Bethe and Heitler \cite{Bethe-Heitler,Heitler1954} based on the
two dominant Feynman diagrams, and similarly the differential cross
section for fully polarized photons
by \cite{BerlinMadansky1950,May1951}\footnote{Misprint corrected
 in \cite{jau}.}.

\begin{itemize}
\item
Only the linear polarization of the incoming radiation takes part in
these 1rst-order Born approximation expressions;
\item
The circular polarization, which was extensively discussed during the
conference, does not.
\item 
The value of the polarization asymmetry, $A$, is close to 0.2 over
most of of the energy range, with low- and high-energy asymptotes of
$\pi/4$
\cite{Gros:2016dmp} and $1/7$.
\item
As the final state is determined by five variables, the definition of
``the'' azimuthal angle of the event can be done in several ways:
examination of the precision of the measurement shows that the optimal
choice is the azimuthal angle of the bisectrix of the direction of the
electron and of the positron
\cite{Gros:2016dmp}.
\end{itemize}

\section{The HARPO project}

A time projection chamber (TPC) is a volume of matter immersed into an
electric field, so that the ionisation electrons produced by the
passage of high-energy charged particles drift and are collected on an
anode plane \cite{Attie:2009zz}.
The anode is segmented so as to provide a 2D image of the electrons
raining on it as a function of drift time.
The measurement of the drift time provides the third coordinate:
An effective way to obtain a fine 3D image of each event.
Here the TPC is used as an active target, that is at the same time the
converter in which the $\gamma$-ray converts and the tracker in which
the two lepton trajectories are measured.

\begin{figure}[htb]
 \includegraphics[width=0.4\linewidth]{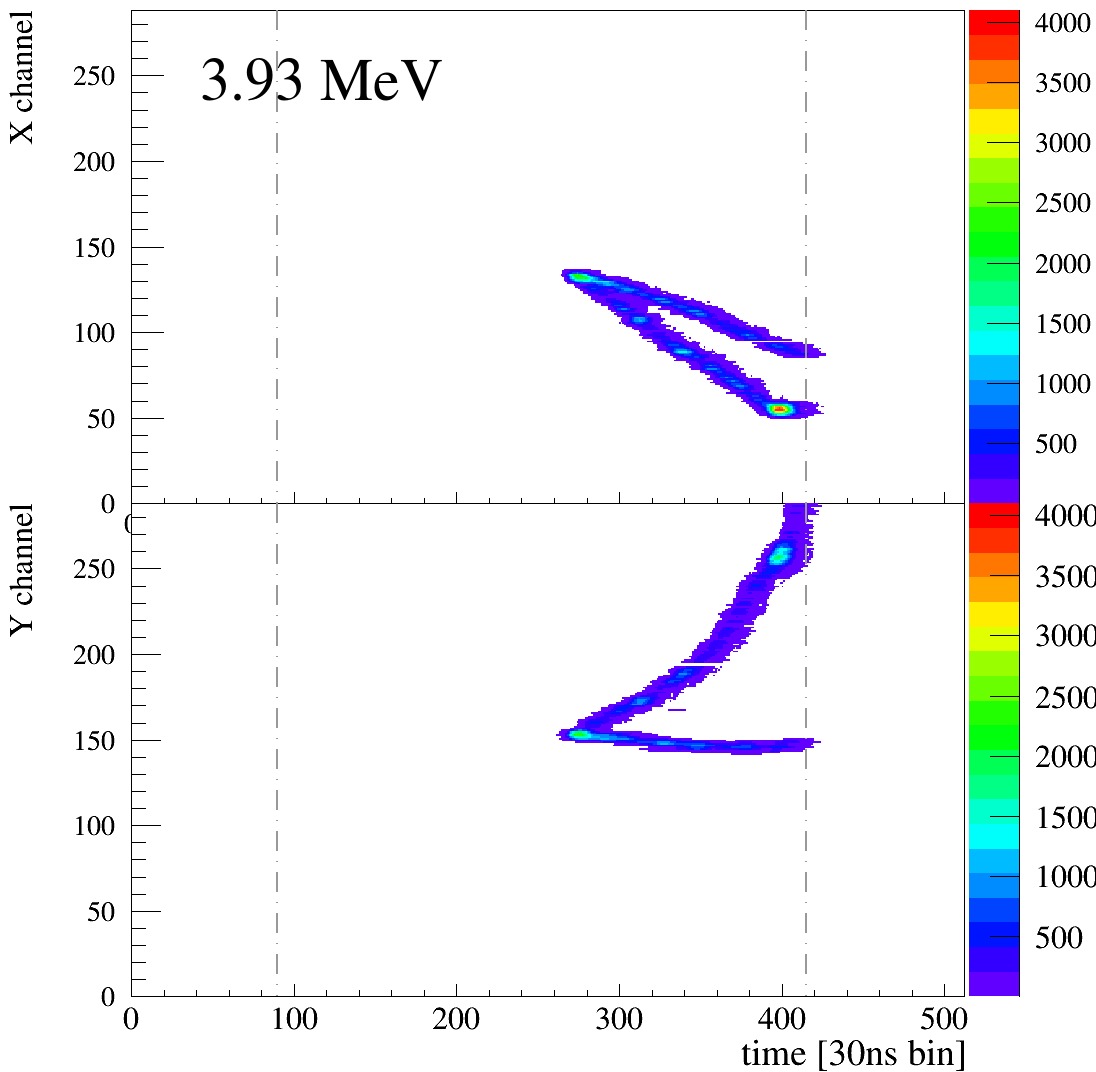}
\hfill
 \includegraphics[width=0.54\linewidth]{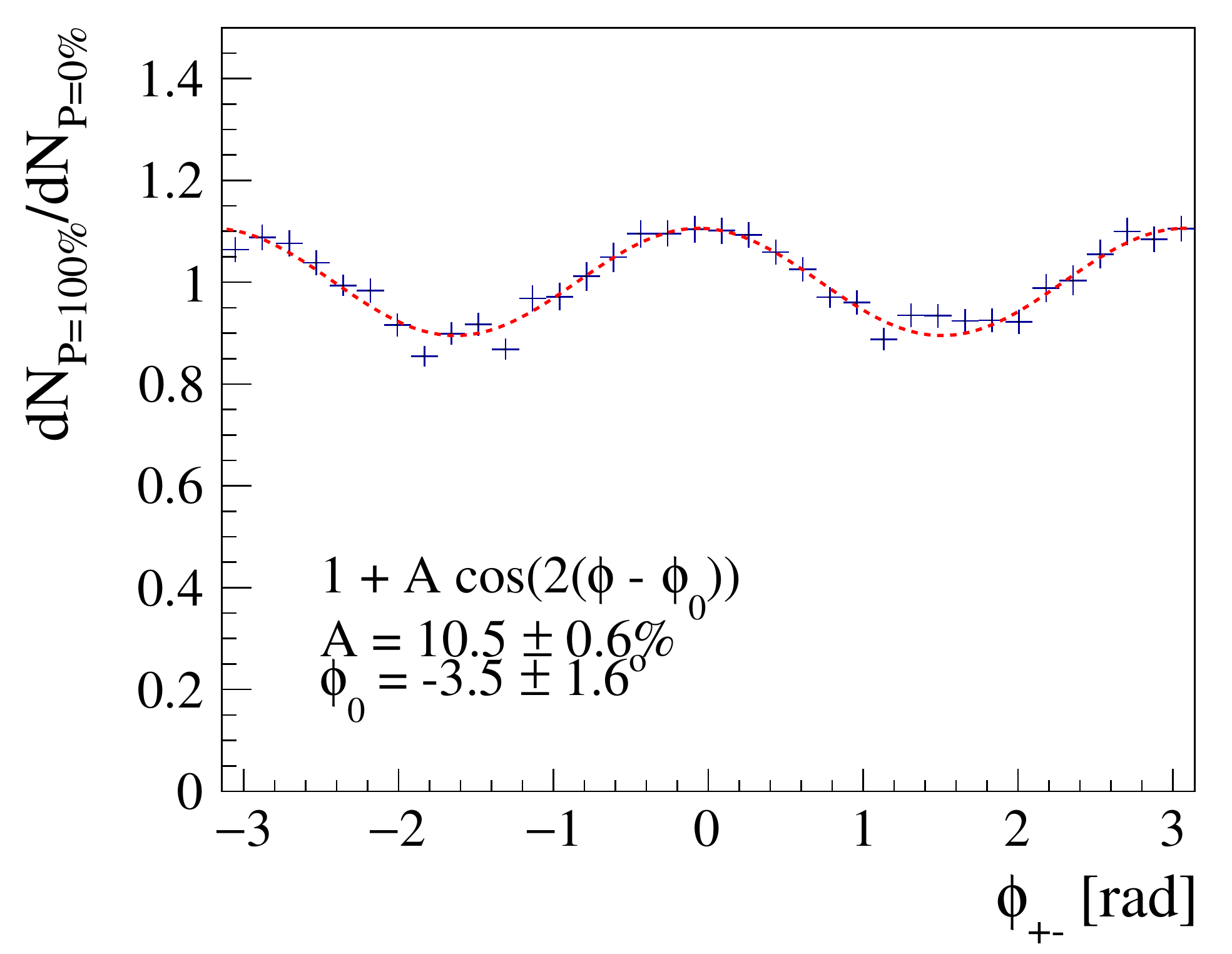}
 \caption{Left: the two ``maps'', that is, the two $x,t$ and $y,t$ projections of a conversion event of a 3.93\,MeV $\gamma$-ray converting to an $e^+e^-$ pair in the 2.1\,bar argon-isobutane (95-5\,\%) gas of the HARPO TPC prototype.
 $x$ and $y$ are the two directions transverse to the drift direction in the TPC; $t$ is the drift time.
 (Right):
 distribution of the azimuthal angle of 11.8\,MeV $\gamma$-rays (ratio of the fully linearly polarized to the linearly non-polarized) converting to an $e^+e^-$ pair in the 2.1\,bar argon-isobutane (95-5\,\%) gas of the HARPO TPC prototype \cite{Gros:2017wyj}.
\label{fig:HARPO}
 }
\end{figure}

In our case the segmentation consists of 2 series of orthogonal strips
along $x$ and $y$, which enables a number of electronics channels that
scales as $2 \times n$ so as to fit into the limited electrical power
available on a space mission (a pad-based system would scale as
$n^2$).
This reduction comes at the cost of a track-assignment ambiguity in
multi-track events.
This issue is easily solved thanks to the wild variation of the energy
deposition along each track: track matching is performed by comparing
the deposited-charge time profiles of the tracks
\cite{Bernard:2012jy,Gros:SPIE:2016}.
It's clear from Fig. \ref{fig:HARPO} left, that shows the $(x,t)$ and
$(y,t)$ signal ``maps'' of a $\gamma$-ray pair-conversion event in our
detector, that the large local ``blob'' close to $t=380\,$ns, almost a
delta-ray, alone, enables an unambiguous track matching.

We designed, constructed, commissioned a gas time projection
chamber (TPC) prototype
\cite{Bernard:2014kwa}
and we exposed it to a $\gamma$-ray beam
provided by the BL01 line of the NewSUBARU facility (U. of Hyogo,
Japan) \cite{Delbart:2015rmp}.
Photons are produced by the inverse Compton scattering of a
laser beam on the electron beam of the 1\,GeV storage ring
\cite{Horikawa2010209}.
By varying the energy of the electron beam and/or the wavelength of
the laser, we could vary the energy of the $\gamma$ rays between 1.7
and 74\,MeV \cite{Delbart:2015rmp}.
The Compton edge of the laser inverse Compton scattering spectrum,
that is, the high part of the $\gamma$-ray energy spectrum, was
selected by collimation on axis.
After collimation, the polarization of the laser beam is almost
entirely transferred to the $\gamma$-ray beam \cite{Sun:2011es}.
Triggering was performed by a dedicated system 
\cite{Geerebaert:2016dyv}.

We developed a Geant4 simulation of the experiment, with TPC
parameters that we have carefully calibrated by comparison with the
properties of the experimental data \cite{Gros:TPC:2016}.
As we found no appropriate $\gamma$-conversion event generator
available on the market, we wrote our own, exact, free from any
high-energy approximation, energy-momentum-conserving, valid down to
threshold, fully 5D and polarized, that we carefully validated
``against'' all 1D analytical expressions that we could find in the
literature \cite{Bernard:2013jea,Gros:2016zst}.
We have demonstrated for the first time in the sub-GeV energy range
in which most of the statistics lies for cosmic sources, a
high-performance polarimetry with an excellent dilution factor
(Fig. \ref{fig:HARPO} and \cite{Gros:2017wyj}).

We also performed a number of hardware developments, such as the
assessment of the long-term quality conservation of the TPC gas in a
sealed mode \cite{Frotin:2015mir}.
We also designed and characterized a prototype series ASTRE
\cite{Baudin}, an upgraded version of the readout chip AGET
\cite{AGET} that includes an improvement of the radiation hardness
from a threshold linear energy transfer (LET) of 3\,MeV/(mg.cm$^2$) to
$\approx$\,20\,MeV/(mg.cm$^2$).
These chips include a self-trigger facility and provide real-time
information of the channels that have seen signal while the drifing
electrons from the TPC volume are raining on the collecting strips: a
space-grade autonomous TPC trigger system is being studied.

\section{Gamma-ray astronomy with an autonoumous active target:
 Optimal measurement of charged particle momentum from multiple scattering with a Bayesian analysis of filtering innovations }

The lower-density space telescopes that are considered these days will imply
 large volume systems so as to maximize the effective area.
The measurement of the photon energy in space telescopes can be
achieved classically either by calorimetry, by the measurement of the
track momenta by magnetic spectrometry or by transition radiation
detection, all systems that would be a challenge to the mass budget on
a space mission.

\begin{figure}[htb]
\includegraphics[width=0.475\linewidth]{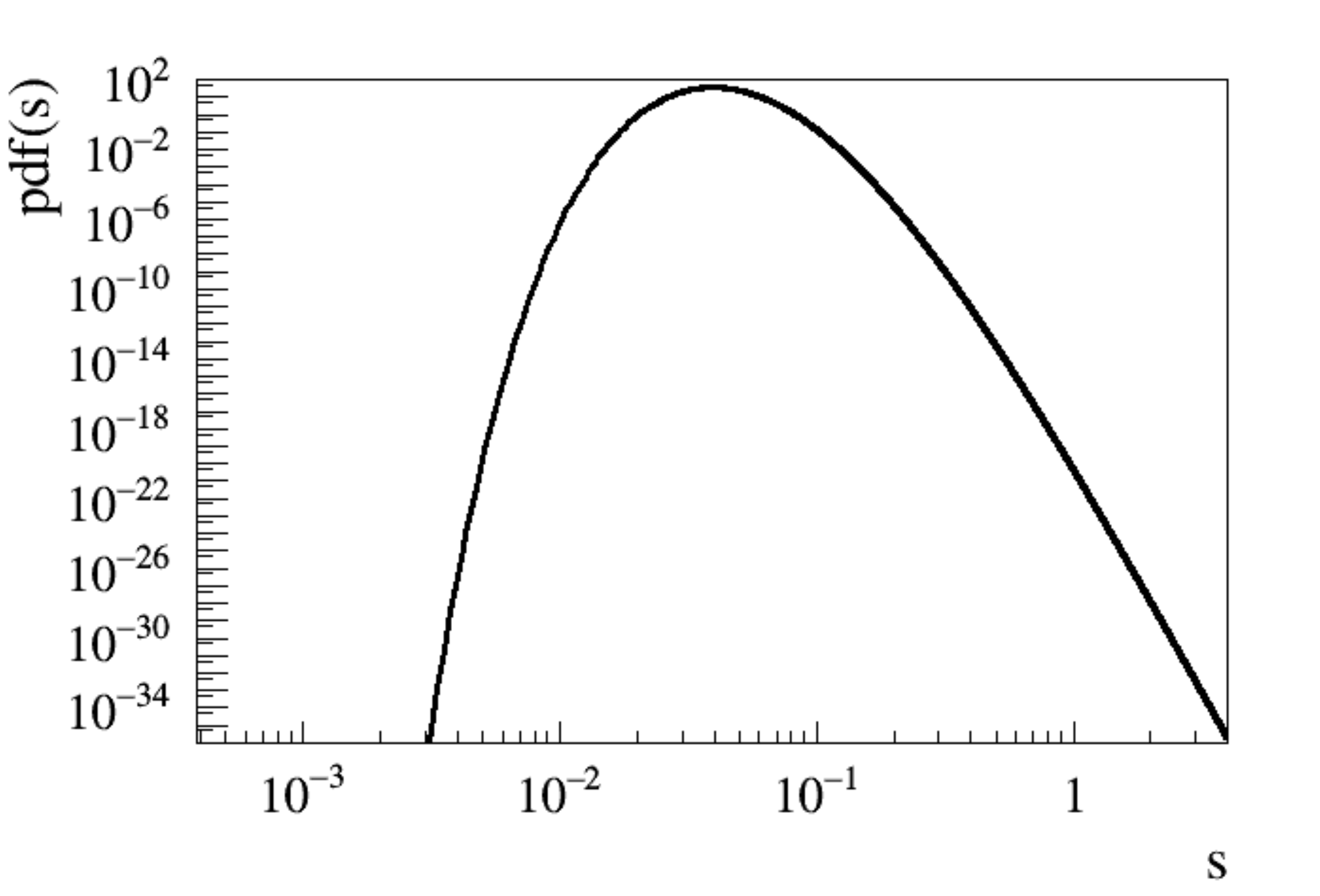}
 \hfill 
 \includegraphics[width=0.482\linewidth]{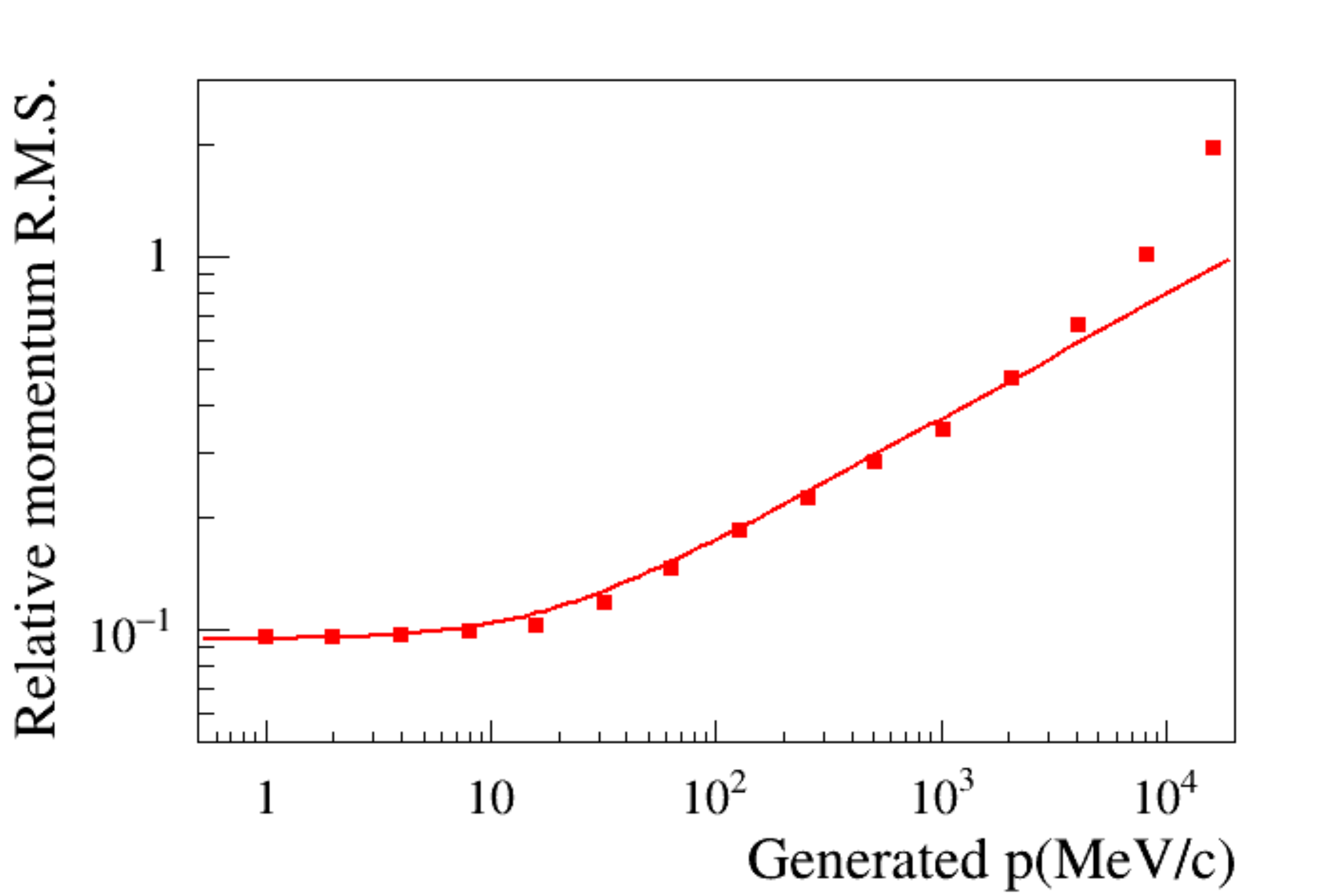}
\caption{Left: Bayesian probability density function as a function of
 $s$, the average multiple-scattering angle variance per unit track
 length, for a 50\,MeV$/c$ track in a silicon
 detector \cite{Frosini:2017ftq}.
On that track, the momentum is measured to be equal to
49.9\,MeV$/c$.
Right: the relative R.M.S of the measured momentum, as a function of
the true track momentum, for a silicon
detector (the curve is the parametrization of eq. (58) of \cite{Frosini:2017ftq}).
\label{fig:momentum}
}
\end{figure}

The momentum of a track can also be measured in the tracker itself, making
use of the dependence of the RMS multiple scattering angle
proportional to the track momentum $\theta_0 \propto 1/p$.
Classically, this is performed by segmenting the track into tracklets,
and measuring the track angle in each tracklet: in that way multiple
angle deflections can be measured along the track \cite{Moliere}.
We had determined what is the tracklet length that optimizes the
momentum precision of that segment method, to find that it depends on
the track momentum .. which is not known yet \cite{Bernard:2012uf}.
Iteration is to be considered.

Recently we developed a segment-free, optimal, unbiased method based
on the Bayesian analysis of the filtering innovations of a series of
Kalman filtered (indexed by the putative track momentum)
\cite{Frosini:2017ftq}.
For a typical silicon wafer stack telescope \cite{E-Astrogam:2016},
the method is expected to be usable up to a couple of GeV/$c$
(fig. \ref{fig:momentum}).

\section{Conclusion}

We have characterized quantitatively the various contributions to the
single-photon angular resolution of pair telescopes and the
improvement that can be expected by the use of lower-density active
targets, we have shown that with gas-detectors the polarimetry of
linearly polarized $\gamma$-rays in the range MeV-GeV is possible
before multiple scattering ruins the polarimetric information.
We
have validated experimentally these results by the characterization of
a prototype on beam.

We have also developed a number of techniques that will be of interest
to make the best use of such telescopes, including an optimal
measurement of track momentum in the tracker itself that will make the
active target an autonomous, low-mass budget space telescope.

\section{Acknowledgments}

I warmfully acknowledge the support of the French National Research
Agency (ANR-13-BS05-0002).

\end{document}